*Predictions of 2019-nCoV Transmission Ending via Comprehensive Methods*

**Tianyu ZENG, Yunong ZHANG, Zhenyu LI, Xiao LIU, Binbin QIU**
School of Data and Computer Science, Sun Yat-sen University, Guangzhou 510006, China
Research Institute of Sun Yat-sen University in Shenzhen, Shenzhen 518057, China

**Summary**
Since the SARS outbreak in 2003, a lot of predictive epidemiological models have been proposed. At the end of 2019, a novel coronavirus, termed as 2019-nCoV, has broken out and is propagating in China and the world. Here we propose a multi-model ordinary differential equation set neural network (MMODEs-NN) and model-free methods to predict the interprovincial transmissions in mainland China, especially those from Hubei Province. Compared with the previously proposed epidemiological models, the proposed network can simulate the transportations with the ODEs activation method, while the model-free methods based on the sigmoid function, Gaussian function, and Poisson distribution are linear and fast to generate reasonable predictions. According to the numerical experiments and the realities, the special policies for controlling the disease are successful in some provinces, and the transmission of the epidemic, whose outbreak time is close to the beginning of China Spring Festival travel rush, is more likely to decelerate before February 18 and to end before April 2020. The proposed mathematical and artificial intelligence methods can give consistent and reasonable predictions of the 2019-nCoV ending. We anticipate our work to be a starting point for comprehensive prediction researches of the 2019-nCoV.

**Article**
Coronavirus, a kind of mammalian avian virus, has caused thousands of infected cases. Two of the most famous coronavirus public health events are the severe acute respiratory syndrome (SARS) and the Middle East respiratory syndrome (MERS), which outbreak in 2003 and 2015, respectively. At the end of 2019, another coronavirus[1], termed as 2019-nCoV, was discovered and began transmission in Wuhan City, Hubei Province, China. Because the outbreak position is one of the most important transportation hubs[2] and the date is close to the annual Spring Festival travel rush[2] (SFTR), the transmission speed of the virus is faster than that of the SARS in 2003. Additionally, due to the fast development of Chinese transportation facilities construction, the disease can spread interprovincially or even internationally, which makes the people and governments concerned.

Due to its high contagion[3, 4], the virus is still spreading fast, and the amounts of patients are still increasing. Thus, the people and governments need to predict the trend of the epidemic and make better decisions to control the transmission. In previous research works, several mathematical models[5] are proposed, and the ordinary differential equation set (ODEs) driven models have the best interpretability and share the largest ratio. In 1927, Kermack and

McKendrick proposed the susceptible-recovered (SR) model[6], which can give the dynamic results of the transmission according to the medical information. Later, the extended models, such as the susceptible-exposed-infectious-recovered (SEIR) model[7], are proposed. However, these models may give inconsistent predictions because of the hypotheses and reality variety. With neural networks[8-11], researchers can generate predictions with the training data[9], and the artificial intelligence (AI) methodology has deeply changed the modeling methods but avoiding the overfitting problems is another important topic.

In this work, we present three model-free predictive methods and propose an ODE combined predictive neural network, multi-model ODEs neural network (MMODEs-NN), to predict the ending of the transmission. To lay a basis for further discussion, we first get the 2019-nCoV daily transmission data[12] from the Chinese Center for Disease Control and Prevention (CDC) and the CDCs of each province in mainland China from January 10 to February 5, 2020. The data contain the confirmed infected amount, recovered amount and death amount. Then, we also collect the population, density and transportation data of each province during the Chinese SFTR[2]. The three model-free methods, which are based on the sigmoid function[10, 14], Gaussian function, and Poisson distribution[15], respectively, only require the daily data. The sigmoid function method requires the total amount of the infected patients, whereas the other two methods require the daily new confirmed data. The MMODEs-NN is a feedforward neural network[16], which takes all the gathered data as the factors and uses the last 4 days data as the test set. The neurons in this neural network are activated by the proposed susceptible-exposed-infected-recovered-susceptible-death (SEIRSD) ODEs, and each of them stores the variable states. Similar to the recurrent neural networks[17], the network can calculate the states for each time step recurrently, and the neuron links can simulate the interprovincial disease transmission in neuron wide propagation and population change according to the transportation data. With certain error evaluation and optimization methods[18], the parameters of the model, such as the disease spreading parameters, can be learned. For more details on the methods and data, please refer to the Method and Supplementary Information sections.

Here we present the prediction results of the 2019-nCoV transmission. Let us denote January 10, 2020, as the 1st day in the training and prediction intervals. When using model-free methods, the results are dependent on the original data. The predictions of these methods are shown in Figs. 1-3, respectively. From these results, one can see that the earliest ending would be the end of February. For the MMODEs-NN, the results are less dependent on the data, so we can train under hypotheses to avoid overfitting and data problems. In general, one has known that the central government of China has taken stronger policies than those in 2003, and we assume that the CDCs in China can count the confirmed patients with no delay. With the help of conjugate gradient method on optimizing the parameters of the disease and transportations, the simulation shows that the ratios between the contact population and the population-area density in other provinces are $8 \times 10^{-4}$, while the ratio in Hubei Province is 0.086 (108x). In this situation, the countrywide total confirmed count would be around 40000. The prediction results are depicted in Fig. 4 and imply that the slowdown would begin on February 7, and the ending would exist before March. However, every disease has a certain time for the researchers to identify. From the latest data, one can see that the data of Hubei Province's total confirmed amount has a boom since January 28, which means the possible existence of this delay. Figs. 5 and 6 suggest the delay situation. The ratio in Hubei Province would be 0.103 (129x), and the countrywide confirmed count would be at the scale of 46000 if the statistical delay is 1 day, while the ratio of Hubei Province would be 0.135 (169x) and the countrywide confirmed count would be at the scale of 58000 if a 2-day delay exists. All the prediction results are listed in Table S7. In these situations, overestimations are acceptable and common in the beginning, and the deceleration date of the transmission would be around February 10, and the ending would happen in March.

What is worth noticing is that the government of Wuhan City declared policies to close the transportation exits of the city[19] on January 23, and later, some other cities in China also announced the similar statements to decelerate the transmission speed. The central government of China also declared other policies to help people and find potential patients under the difficulties, such as prolonging the Spring Festival vacation and conducting strict body temperature measurements in blocks[19]. We consider these policies and code them with the proposed network. According to transportation data in the Supplementary Information, the traffic during the 2020 Spring Festival in China decreases by 90% compared with that in 2019. These policies help to control the disease and shrink the total confirmed amounts in each province. We also conduct another simulation without these policies, and the results plotted in Fig. 7 show that the final infected patient amount would reach 450000. In this situation, the most optimistic deceleration date is February 26, and the ending would be around April 28. Thus, these policies work, and their effects are undoubtedly obvious.

In summary, the above numerical prediction results give the expectations on the ending of the transmission. From the experiments, one can see that the model-free methods and the proposed MMODEs-NN can give consistent and confident predictions on the disease and the transmission of the coronavirus would be under control soon. However, every predictive method has its corresponding advantages and shortcomings. The model-free methods show statistical predictions by learning the general form of the developing data, while the proposed MMODEs-NN generates predictions from the simulation by synthesizing the data independency of the ODEs dynamic system and the fitting power of neural networks. For model-free methods, due to the dependency of the data, they are fast to learn the data pattern but would be more likely to make errors if the real-time data are incorrect or have latencies; for the ODEs-NN combined model, it requires more time to train and generate overfitting problems but it can find the human-like errors in the training data and provide more independent predictions. Both of them have strengths, and the predictions would be more robust and more possible to happen if their results are consistent. Additionally, a variety of uncertain random events happen in real life. For example, the boom in Zhejiang is related to clustering and weak related to the concerned factors in the simulations of MMODEs-NN. However, for the total confirmed amount, our models are relatively accurate, even if the CDC has lowered the standard of judging the confirmed patients since February 9. Thus, we summarize the predicted endings in Table 1 and conservatively estimate that the transmission of the 2019-nCoV would slow down before February 18 and the transmission of this year would finally come to the ending before April. For more detailed data, please refer to the Supplementary Information.

In conclusion, this study has presented comprehensive predictions for the 2019-nCoV transmission ending, providing a guideline for policymaking and a panorama view on the disease future development. Besides the ending, the long treatment cycle, the potential multi-cycle transmission shown in Fig. 1, and the potential high death rate shown in Figs. 4-7 are also worth and important for the medical staff to care about. Based on this work, we expect to do more and deeper predictive researches on the geographical spreading problems with model-free methods and the proposed SEIRSD MMODEs-NN. For it is the first time to apply the comprehensive methods to predict the 2019-nCoV transmission procedure, we anticipate our work to be a starting point for prediction researches of the 2019-nCoV.

**Table 1. Final predictions of the 2019-nCoV transmission endings via model-free methods and SEIRSD MMODEs-NN**

| Method | Sigmoid | Gaussian | Poisson | NN | NN with 1-day delay | NN with 2-day delay | NN without limitation policies |
|---|---|---|---|---|---|---|---|
| **Deceleration** | Feb. 4 | Feb. 11 | Feb. 6 | Feb. 7 | Feb. 9 | Feb. 10 | Feb. 26 |
| **Ending Date** | Feb. 28 | Mar. 10 | Feb. 29 | Feb. 24 | Feb. 28 | Mar. 3 | Apr. 28 |

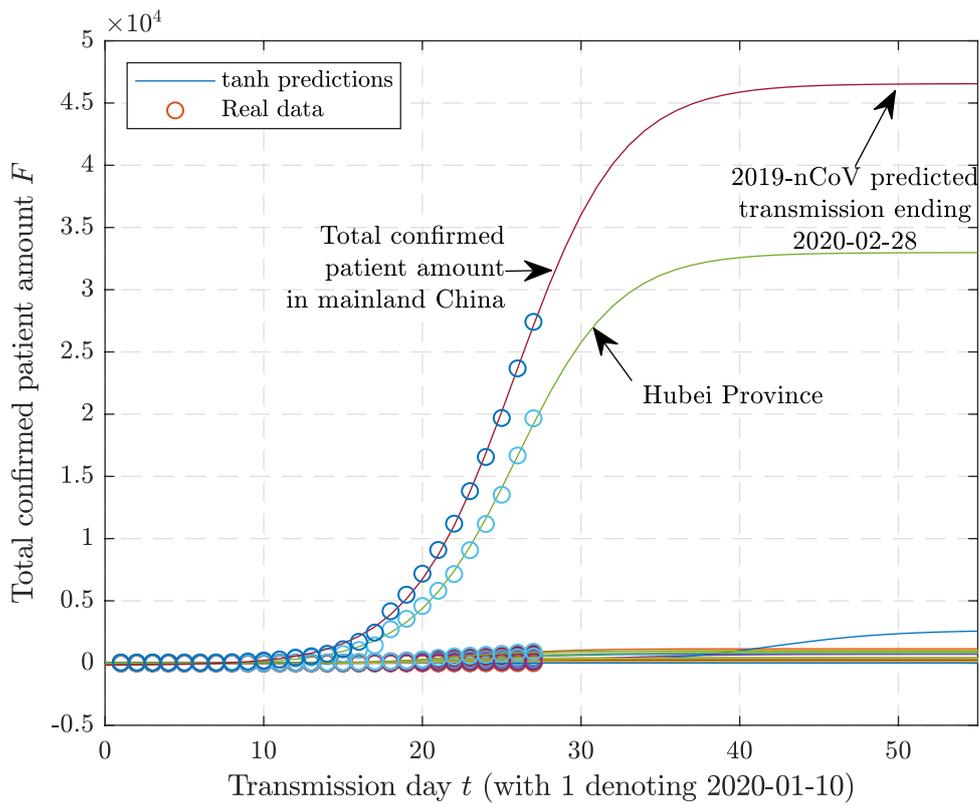

**Fig. 1 Predictions of 2019-nCoV transmission ending in mainland China via sigmoid functions.** To reduce the count of the function parameters, we adopt the tanh function to fit the data. One can see from the fitting results that the potential transmission ending of the 2019-nCoV would be February 28, 2020, and the final total confirmed amount would be 46000.

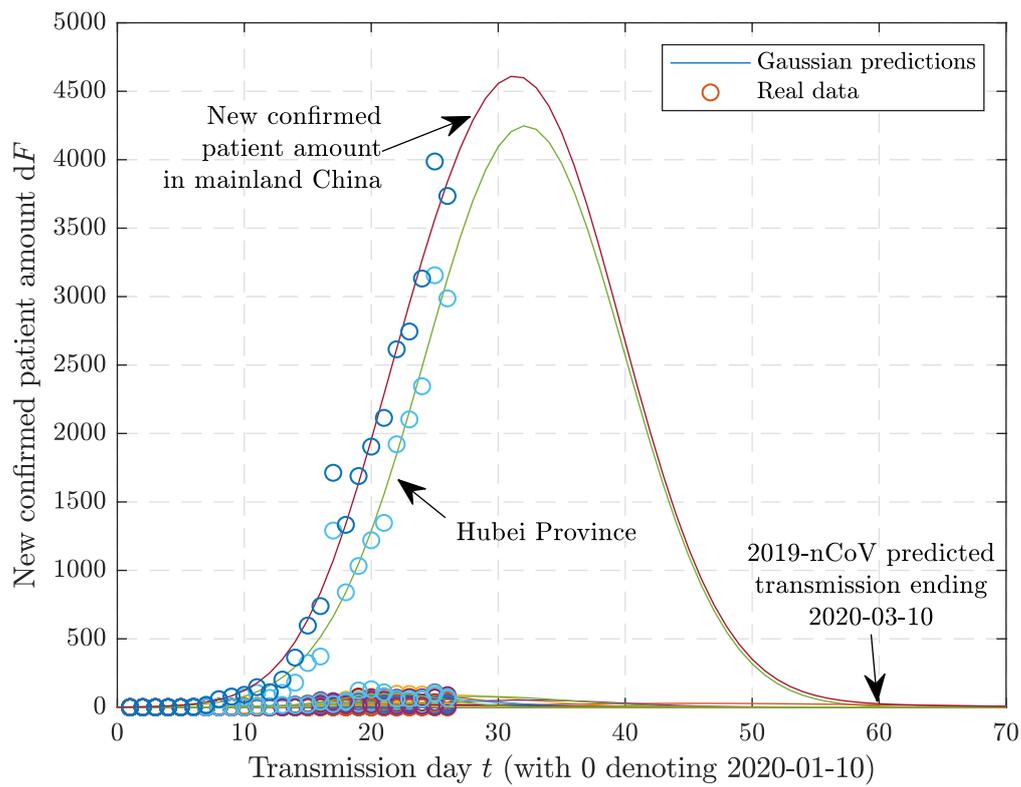

**Fig. 2 Predictions of 2019-nCoV transmission ending in mainland China via Gaussian functions.** We take the Gaussian function as the target function type. By using new confirmed data, the fitting results can be obtained and indicate that the potential transmission ending of the 2019-nCoV would be March 10, 2020.

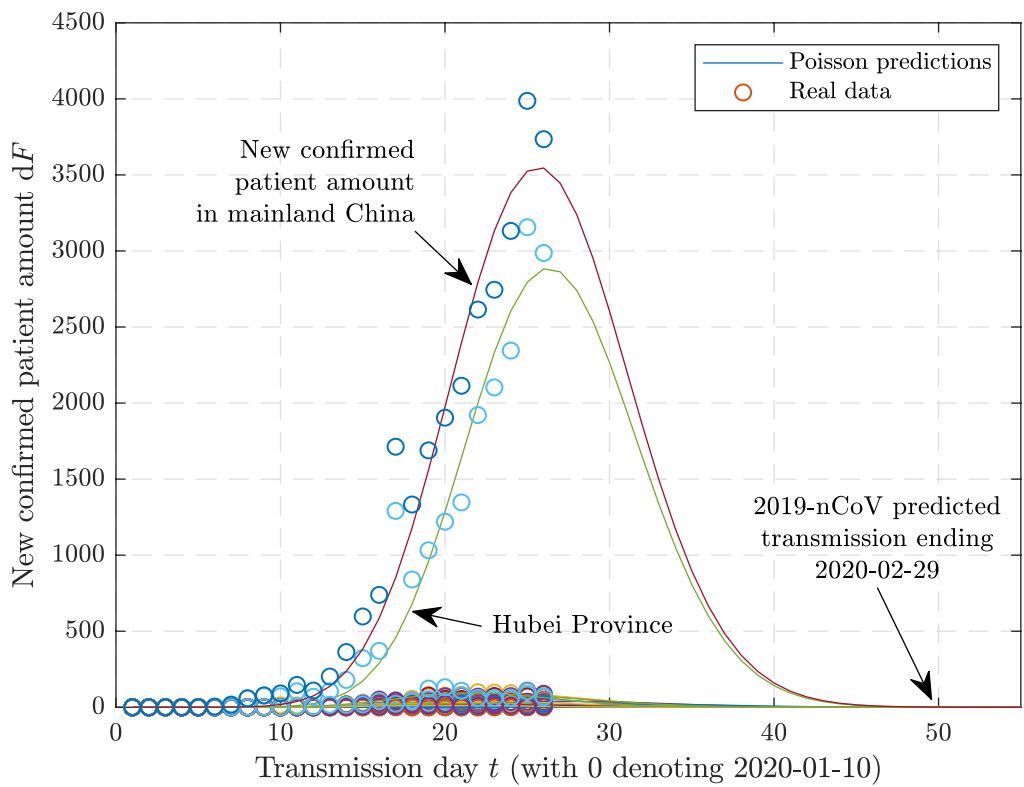

**Fig. 3 Predictions of 2019-nCoV transmission ending in mainland China via Poisson distributions.** Considering the properties of the Poisson distribution, we construct the continuous Poisson function and fit the provincial data. The fitting results can be obtained and indicate that the potential transmission ending of the 2019-nCoV would be February 29, 2020.

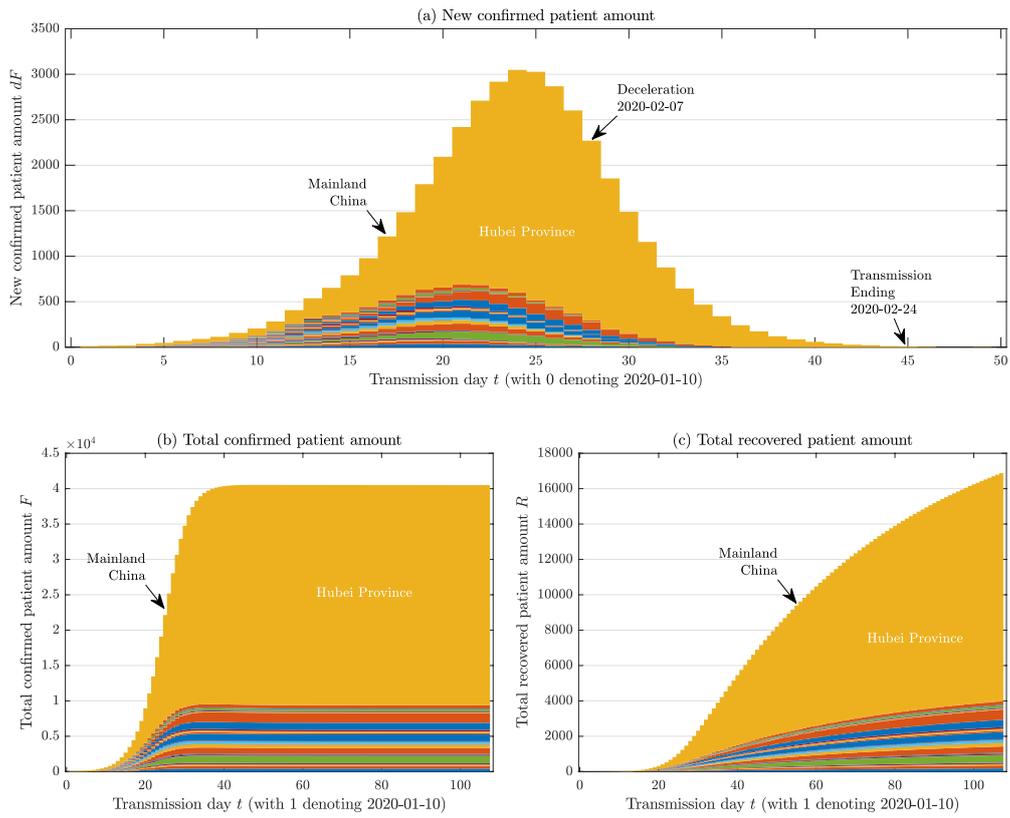

**Fig. 4 Predictions of 2019-nCoV transmission ending in mainland China via SEIRSD MMODEs-NN.** By using the proposed MMODEs-NN activated by SEIRSD ODEs, we can train the data from January 10 to February 1 and test with the data in the last 4 days. The results in (a) show the potential deceleration would begin on February 7, and the ending would be on February 24, the results in (b) show the potential total confirmed patient amount are at the scale of 40000, and the results in (c) show the recovered patient amount.

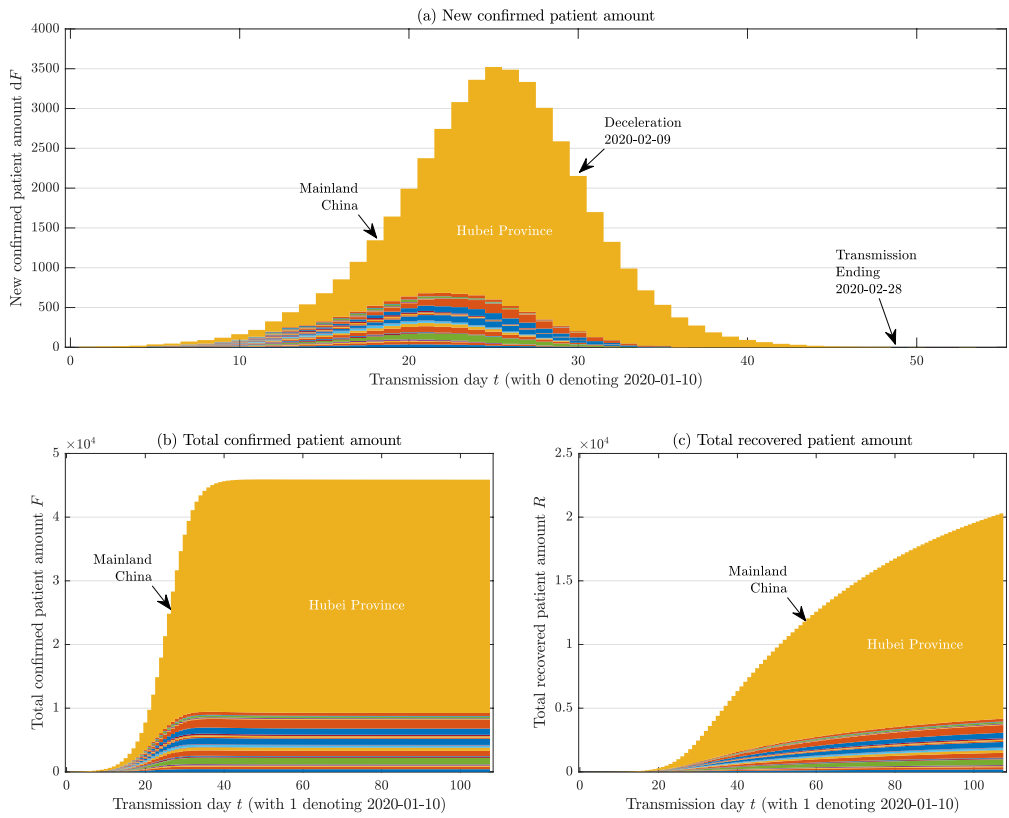

**Fig. 5 Predictions of 2019-nCoV transmission ending in mainland China via SEIRSD MMODEs-NN under 1-day statistical delay in Hubei Province.** The results in (a) show the deceleration would begin on February 9, and the ending would be on February 28, while (b) shows the potential total confirmed amount would be 46000.

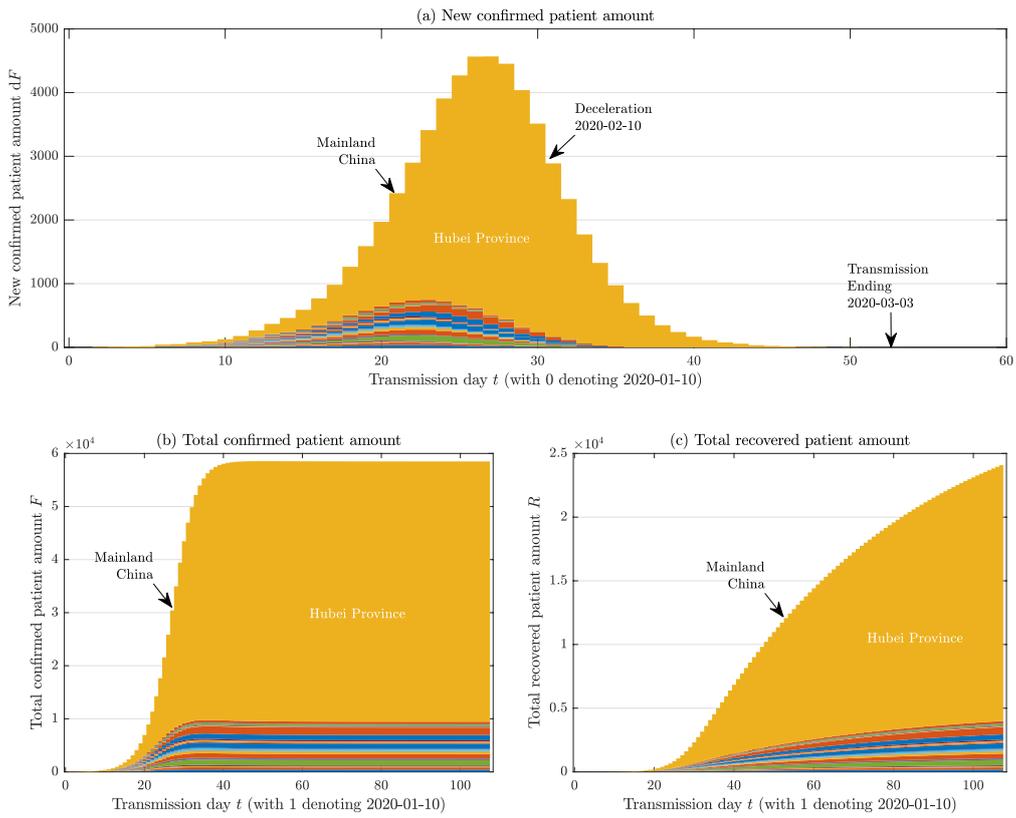

**Fig. 6 Predictions of 2019-nCoV transmission ending in mainland China via SEIRSD MMODEs-NN under 2-day statistical delay in Hubei Province.** The results in (a) show the deceleration would begin on February 10, and the ending would be on March 3, and figure (b) shows the potential total confirmed amount would be around 58000.

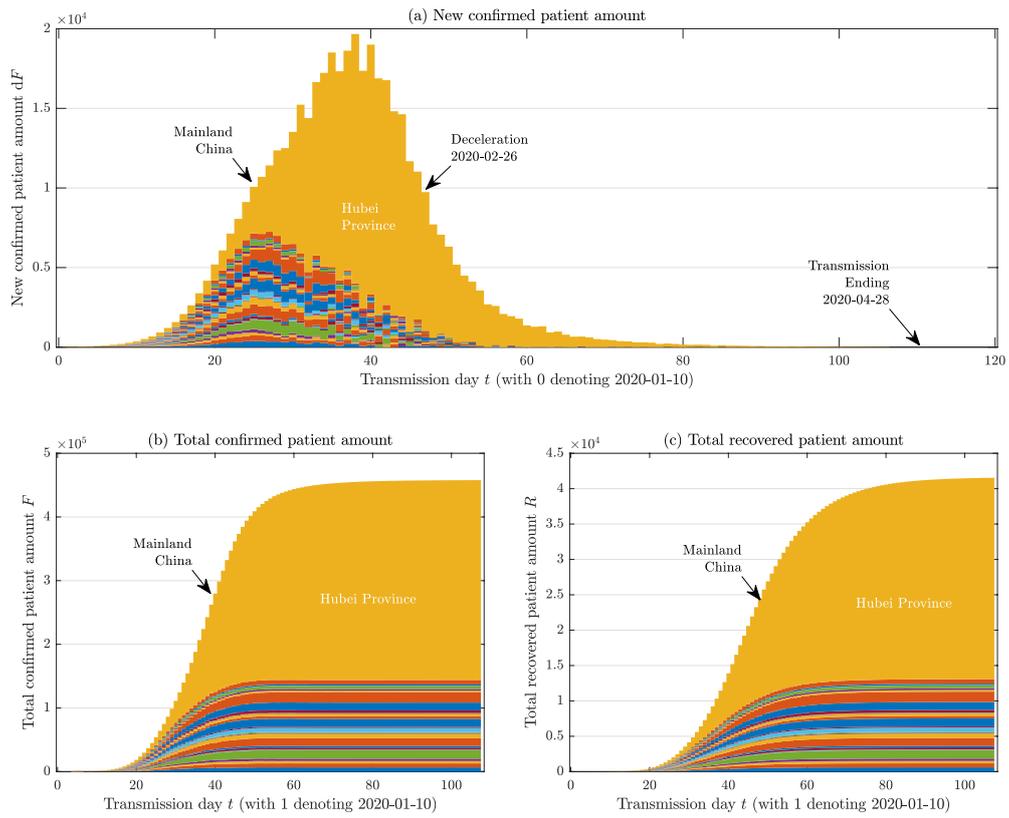

**Fig. 7 Predictions of 2019-nCoV transmission ending in mainland China via SEIRSD MMODEs-NN without transportation limitation policies.** The results in (a) show the deceleration would begin on February 26, and the ending would be on April 28 in this situation, figure (b) shows the potential total confirmed amount would be around 450000.

# Methods

## I. Model-Free Methods

In this study, we present three model-free methods to predict the 2019-nCoV transmission ending. They are based on the sigmoid function, Gaussian function, and Poisson distribution, respectively. Here we present their technical details.

### A) Sigmoid Function

Sigmoid functions, belonging to a class of S-shaped functions[10, 14], include the sigmoid function and tanh function. In the epidemiological models, such as SR and SI models, the results are sigmoid-like functions. In this work, we adopt the tanh function to fit the provincial data. The definition of tanh function is

$$f_1(t, a, b, c, d) = a \tanh(b(t + c)) + d,$$

where $t$ denotes the time, the parameters $a$, $b$, $c$, and $d$ refer to the amplification, scale, initial phase, bias, respectively. By using the confirmed data in Table S1, we can get the fitting results for each province. The countrywide confirmed predictions are shown in Fig. 1, and the detailed data are depicted in Table S2.

### B) Gaussian Function

Gaussian distribution is a common existed distribution in real life. From the aspect of time, it means that the probabilities of a certain-theme event will follow a corresponding Gaussian distribution in continuous time. In this work, we assume the probability of an infected patient going to the hospital follows a Gaussian function, which can be written as

$$f_2(t, a, b, c) = a \exp\left(-\left(\frac{x-b}{2c}\right)^2\right),$$

where $t$ denotes the time, $a$ denotes the amplification, $b$ denotes the expectation, and $c$ denotes the standard deviation of the Gaussian distribution. By using the daily newly confirmed data in Table S1, we can get the ending date for each province, and the prediction results are plotted in Fig. 2. The detailed data are depicted in Table S3.

### C) Poisson Distribution

Poisson distribution is frequently used as random events distribution[15]. In this work, we combine the Poisson distribution with Stirling's approximation[20] to get the fitting parameters. The Poisson distribution fitting function is

$$f_3(t, a, b) = \frac{ab^t \exp(t-b)}{t^t \sqrt{2\pi t}},$$

where $t$ denotes the time, $a$ denotes the amplification, and $b$ denotes the expectation in a unit time step. By using the daily newly confirmed data in Table S1, we can get the ending date for each province, and the prediction results are plotted in Fig. 3. The detailed data are depicted in Table S4.

## II. SEIRSD ODEs and Multi-Model ODEs Neural Network

The SEIRSD denotes the susceptible-exposed-infected-recovered-susceptible-dead loop. A patient may experience a susceptible period and an exposed period and would be confirmed to be infected. The patient would have a chance to recover and a chance to become dead. The recovered patient might have another chance to get in the loop for another time. For the patient, who is in the exposed state, he/she may become recovered at a certain chance. To describe the loop in a more comprehensive way, here we give the ordinary differential equation set of the SEIRSD, that is,

$$\frac{dS}{dt} = -\beta IS + \theta R, \quad \frac{dE}{dt} = \beta IS - (\alpha + \gamma_1)E, \quad \frac{dI}{dt} = \alpha E - (\gamma_2 + \delta)I, \quad \frac{dR}{dt} = \gamma_1 E + \gamma_2 I - \theta R, \quad \frac{dD}{dt} = \delta I,$$

where the capitalized variables mean the amount of the corresponding words, while the parameters $\alpha, \beta, \gamma_1, \gamma_2, \delta$, and $\theta$ denotes the confirmed rate, the transmission rate, the recovery rate of the exposed patients, the recovery rate of the confirmed patients, the death rate, and the recurrence rate, respectively. Like the $R_0$ value for virus transmission[21], these parameters are related to the disease properties and can be learned with the MMODEs-NN. In general situations, these parameters would not change rapidly, except for the variation. The hypothesis shared by the ODE-oriented model, which is the total count of the system stays unchanged[6], can be extended[7] by the proposed neural network to be the conservation of the involved people ratio. The collected data are the historical provincial total confirmed amounts, the total recovered amounts, and the death amounts, so another two ordinary differential equations can be added, they are,

$$\frac{dF}{dt} = \alpha E, \quad \frac{dC}{dt} = \gamma_2 I,$$

where $F$ and $C$ denote the provincial total confirmed amount and recovered amount of the transmission, respectively.

The ODE neural network[16] is a kind of time-series neural network. The neuron activation methods of this kind are ODE or ODEs instead of nonlinear activation functions. The neuron can store the states of each time step, and their cell states can be activated and propagate to the cells in the next layer. In this work, we present a fully connected feedforward SEIRSD ODEs-activated neural network with multi-model techniques. To make a robust simulation, the proposed network is designed to have 31 SEIRSD-activated neurons in each layer which represents the corresponding time step, and the structure is shown in Fig. S1. The neuron can store some basic information of the corresponding province, such as the population and density. The links between the layers are fully connected to propagate the interprovincial population change during the SFTR. The weights of the links are controlled by the transportation data, which is listed in Table S5. Besides the SEIRSD model in each neuron, the sub-models that we concerned about are the 2020 SFTR model, the noncontact-contact model, and the statistical delay model. The 2020 SFTR model requires moving-out transportation data. In the last third of January, interprovincial transportations have decreased by 90%, and we formulate it as

$$S_{\text{out}}(t, A, B) = \frac{\mathcal{D}_t(0.9T_{A,B} + 0.2r)S}{100000}, \quad E_{\text{out}}(t, A, B) = \frac{\mathcal{D}_t(0.9T_{A,B} + 0.2r)E}{P_{E\text{out}}}, \quad \mathcal{D}_t = 1 - \frac{P_{\mathcal{D}}}{1 + \exp(-2(t-12))},$$

where $S_{\text{out}}$ and $E_{\text{out}}$ denote the amount of the moving-out population and exposed patients, respectively, while $T$ and $\mathcal{D}$ denote the interprovincial transportation ratios and the restriction force, respectively. For the parameters, $A$ and $B$ denote the departure and arrival province, respectively, $r$ denotes a random value, $P_{E\text{out}}$ denotes the exceed rate of the exposed patients, and $P_{\mathcal{D}}$ denotes the shrink rate of the transportation. From the above, one can see that $P_{\mathcal{D}}=0.9$. The noncontact-contact model means that we assume that there is a part of the population who is hard to get in contact with the potential virus carriers, such as the self-isolated people or the students. The amount of these people can be evaluated with a formula, that is,

$$S_C(t, A) = \frac{P_C \mathcal{D}_t \rho(A) S}{\max \rho},$$

where $S_C$ denotes the virus-contact population, $\rho$ denotes the population density and the parameter $P_C$ denotes contact ratio. The $S_C$ will get involved in the SEIRSD ODEs calculation with other variables in the normalized form. As for the statistical delay model, it is used to calculate the delay of patient statistic work. If some confirmed patients do not be found or a period is needed to research the virus, we take the time duration as the statistical delay. In the SEIRSD MMODEs-NN, thanks to the time-flattened structure, we can keep some states of the neurons without dragging the layer wide propagations.

To simplify the calculations, some of the parameters, such as $P_{\mathcal{D}}$, can be directly obtained from the related statistic works. We assume that Hubei Province might have a statistical delay at the beginning of the 2019-nCoV transmission. Thus, in this model, we have 6 virus-related and 1 sub-model parameters to optimize. The error evaluation methods of the historical provincial total confirmed amount and death amount are the mean absolute errors (MAE). With the help of the conjugate gradient optimizer[18], the parameters can be learned, and the data trend can be fitted. Figs. S2 and S3 show the visualizations of the provincial transmissions in mainland China on February 28 and March 31. For more data, please refer to Table S6 for the training details and Table S7 for the predictions.

**References**

1. Zhou, P., et al. A pneumonia outbreak associated with a new coronavirus of probable bat origin. *Nature*, (2020).
2. Li, J., Ye, Q., Deng, X., Liu, Y., and Liu, Y. Spatial-temporal analysis on Spring Festival travel rush in China based on multisource big data. *Sustainability* **8**, 11, 1184, (2016).
3. Hui, D., et al. The continuing 2019-nCoV epidemic threat of novel coronaviruses to global health—The latest 2019 novel coronavirus outbreak in Wuhan, China. *International Journal of Infectious Diseases* **91**, 264-266, (2020).
4. World Health Organization. Infection prevention and control during health care for probable or confirmed cases of novel coronavirus (nCoV) infection. *Interim Guidance. Geneva: World Health Organization*, (2015).
5. Roberts, F. *Discrete Mathematical Models, with Applications to Social, Biological, and Environmental Problems*, Prentice-Hall Englewood Cliffs, Upper Saddle River, New Jersey, USA, (1976).
6. Kermack, W. and McKendrick, A. A contribution to the mathematical theory of epidemics. *Proceedings of the royal society of London. Series A, Containing papers of a mathematical and physical character* **115**, 772, 700-721, (1927).
7. Li, M. and Muldowney, J. Global stability for the SEIR model in epidemiology. *Mathematical biosciences* **125**, 2, 155-164, (1995).
8. Zhang, Y., Chen, D., and Ye, C. *Toward Deep Neural Networks: WASD Neuronet Models, Algorithms, and Applications*, Chapman and Hall/CRC, Boca Raton, Florida, USA, (2019).
9. Zhang, Y., Guo, D., Luo, Z., Zhai, K., & Tan, H. CP-activated WASD neuronet approach to Asian population prediction with abundant experimental verification. *Neurocomputing* **198**, 48-57, (2016).
10. Zhang, Y., Ding, S., Liu, X., Liu, J., and Mao, M. WASP neuronet activated by bipolar-sigmoid functions and applied to glomerular-filtration-rate estimation. *The 26th Chinese Control and Decision Conference (2014 CCDC)*, 172-177, (2014).
11. Zhang, Y. *Analysis and Design of Recurrent Neural Networks and Their Applications to Control and Robotic Systems*, Ph.D. Thesis, Chinese University of Hong Kong, Hong Kong, (2002).
12. CCDC, *Epidemic Update and Risk Assessment of 2019 Novel Coronavirus*, CCDC, Beijing, China, http://www.chinacdc.cn/yyrdgz/202001/P020200128523354919292.pdf, (in Chinese, 2020).
13. Bol, P. and Ge, J. China historical GIS. *Historical Geography* **33**, 150-152, (2005).
14. Hosmer Jr, D., Lemeshow, S., and Sturdivant, R. *Applied Logistic Regression*, John Wiley & Sons, Hoboken, New Jersey, USA, (2013).
15. Consul, P. and Jain, G. A generalization of the Poisson distribution. *Technometrics* **15**, 4, 791-799, (1973).
16. Zhang, Y., Guo, D., and Li, Z. Common nature of learning between back-propagation and Hopfield-type neural networks for generalized matrix inversion with simplified models. *IEEE Transactions on Neural Networks and Learning Systems* **24**, 4, 579-592, (2013).
17. Williams, R., Hinton, G., and Rumelhart, D. Learning representations by back-propagating errors. *Nature* **323**, 6088, 533-536, (1986).
18. Moller, M. *A Scaled Conjugate Gradient Algorithm for Fast Supervised Learning*, Aarhus University, Aarhus, Denmark, (1990).

19. Nature, Coronavirus latest: infections in China pass 20,000. *Nature News*, https://doi.org/10.1038/d41586-020-00154-w, (2020).
20. Marsaglia, G. and Marsaglia, J. A New Derivation of Stirling's Approximation to *n*!. T*he American Mathematical Monthly* **97**, 9, 826-829, (1990).
21. Liu, T., et al. Transmission dynamics of 2019 novel coronavirus (2019-nCoV). *bioRxiv*, https://www.biorxiv.org/content/10.1101/2020.01.25.919787v1.abstract, (2020).

## *Supplementary Information*

**More About SEIRSD Model, MMODEs-NN, and Simulations**

In the article part, we give a basic view of the SEIRSD model from the ODEs aspect and the corresponding ODEs network. Here we present more detailed illustrations of them.

For the SEIRSD model, Fig. S1(a) displays the structure and the state transition diagram, corresponding to the ODEs of the SEIRSD model that we have shown, which is

$$\frac{dS}{dt} = -\beta IS + \theta R, \quad \frac{dE}{dt} = \beta IS - (\alpha + \gamma_1)E, \quad \frac{dI}{dt} = \alpha E - (\gamma_2 + \delta)I, \quad \frac{dR}{dt} = \gamma_1 E + \gamma_2 I - \theta R, \quad \frac{dD}{dt} = \delta I.$$

For all the variables in the ODEs, we have to change them into the normalized form, and for the parameters, i.e., $\alpha, \beta, \gamma_1, \gamma_2, \delta,$ and $\theta$, they are assigned within [0, 1] corresponding to the rate of their meanings. To simulate the transportation population, we add the inputs and outputs to the susceptible and exposed population[1]. From the meaning of the state transition diagram, one can see that a person, who is involved with the disease, has a procedure and would be a virus carrier if he/she left the province in the exposed state. With neuron wide propagation, we can see the final interprovincial transmission results.

For the SEIRSD MMODEs-NN, one can see from the structure depicted in Fig. S1(b) that the network is similar to the fully connected neural network. To simulate the true environment in the Chinese Spring Festival travel rush (SFTR), we assume that the interprovincial transportation population during the SFTR is large enough[2] to ignore the geographical distances, and the weights of each neuron link are obtained with the transportation data displayed in Table S5. What is worth noticing is that each neuron consists of a SEIRSD sub-model, and the interprovincial links can only propagate the transportation population in the form of counts, while the self-links can propagate the neuron states[3, 4]. For the network initialization, the neurons are loaded with the population and density information of the corresponding province. The network calculates the errors on the training set automatically and adopts the parameters with the best performance on the test set. Similar to the recurrent neural network[4, 5], the network shares the parameters and makes predictions by calculating the neuron states in a layer-layer loop. If we want to simulate the delayed situation, we can first propagate and reset the delayed neuron states, which is similar to remove the self-link of the delayed neuron. With the help of numerical optimization methods, such as conjugate gradient[6] and gradient descendent, the parameters in the network can be learned.

After the numerical simulations, whose results are shown in the article, we believe our network can predict the patient counts with 10% errors. These errors might be aroused by the random events that cannot be predicted, such as clustering events happened in Zhejiang and Guangdong. Thus, simply judging the deceleration from the descendent of the newly confirmed patient count would be more likely to make errors. The standard that we judge the starting date of the transmission deceleration is different from the differential methods. We adopt the first day whose new confirmed patient amount is smaller but 1.22 times larger than the amount on the day before, and larger than 1.22 times the amount on the day after. More specifically, the ratio 1.22 is equal to 11/9, which is consistent with the error bounds of the proposed network. By judging from the prediction values, we can make sure the true deceleration dates without making error predictions.

We also visualize the prediction results with the map of China. Figs. S2 and S3 depict the predictions of the total confirmed amounts on February 28 and March 31. One can see from the results are close to each other, which means that the transmission of the 2019-nCoV would stop before April. Thus, we conservatively estimate that the transmission of the 2019-nCoV would slow down before February 15 and the transmission of this year would finally come to the ending before April.

However, the ending of the transmission does not mean the vanishment of the 2019-nCoV. The meaning of vanishment means that all the patients are recovered or died, and the infected probabilities are lower than the beginning of the outbreak, while the transmission ending means that the transmission is under control, and the total confirmed amount would not increase in a large scale. In general, after the ending, the virus would be less likely to break out again, but still has the variation probability and arouse multi-cycle transmission, which is important for the medical staff to care about. What is more, from the simulations shown in the article, we can see that the treatment cycle is long, and the potential death rate is high. Although the medical levels in the provinces are not considered, the data from January 10 to February 5 imply the information, and the optimized simulations have shown this warning. For a deeper estimation, we conservatively estimate that the final vanishment would be before August 2020.

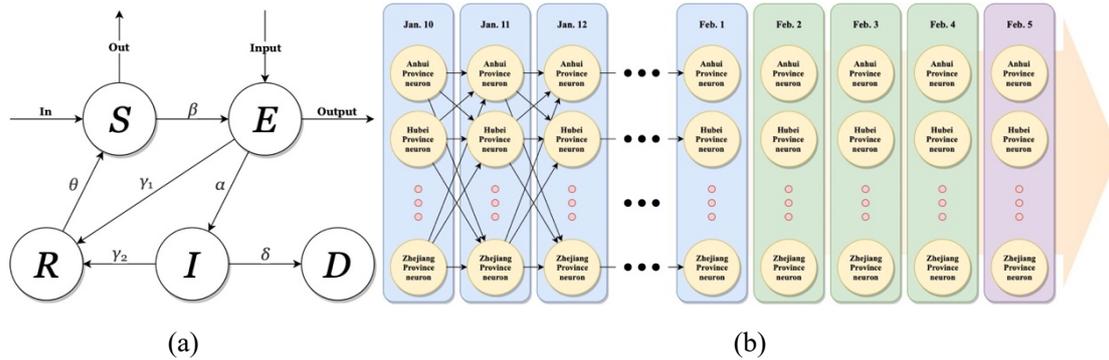

**Fig. S1 (a) SEIRSD neuron structure and state transition diagram. (b) Structure of SEIRSD MMODEs-NN for predicting 2019-nCov transmission ending.** We first gather the provincial basic information to initialize the network. In each neuron, there is a SEIRSD model. The weights are calculated with the transportation data and the 2020 SFTR model. For each time step, the neurons receive the inputs from the last time step neurons, reinitiate the neurons with the latest states, calculate the SEIRSD model with the noncontact-contact model in the current time step, and propagate to next layer. The test set in the simulations are the latest 4-day data, and the error evaluation method is MAE. The parameters in the network can be learned with conjugate gradient and gradient descendent.

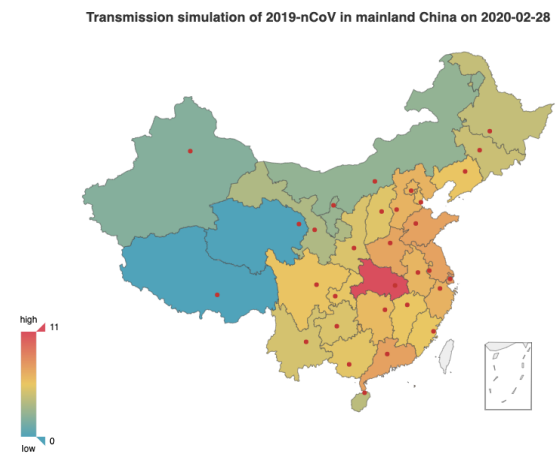

**Fig. S2 Total confirmed amounts visualization of 2019-nCoV transmission simulation in mainland China on February 28.** This figure is log-scaled. One can see that Hubei Province is of the largest patient amount, while Beijing, Shanghai, Guangdong, Zhejiang, Anhui, and Henan are of a large amount. The provinces, such as Qinghai and Tibet, are of small amounts, which is close to reality.

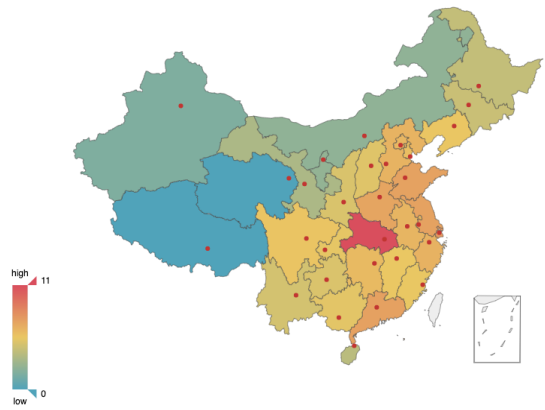

**Fig. S3 Total confirmed amounts visualization of 2019-nCoV transmission simulation in mainland China on March 31.** This figure is log-scaled. One can see that the results are close to the ones in Fig. S2, which means that the transmission of the 2019-nCoV would stop before April.

**Table S1. Historical provincial data of 2019-nCoV confirmed patient amount from January 10 to February 5 (Part 1)**

|     | 10 | 11 | 12 | 13 | 14 | 15 | 16 | 17 | 18  | 19  | 20  | 21  | 22  | 23  |
|-----|----|----|----|----|----|----|----|----|-----|-----|-----|-----|-----|-----|
| AH  | 0  | 0  | 0  | 0  | 0  | 0  | 0  | 0  | 0   | 0   | 0   | 0   | 1   | 9   |
| BJ  | 0  | 0  | 0  | 0  | 0  | 0  | 0  | 0  | 0   | 0   | 5   | 10  | 14  | 26  |
| CQ  | 0  | 0  | 0  | 0  | 0  | 0  | 0  | 0  | 0   | 0   | 0   | 5   | 6   | 9   |
| FJ  | 0  | 0  | 0  | 0  | 0  | 0  | 0  | 0  | 0   | 0   | 0   | 0   | 1   | 5   |
| GD  | 0  | 0  | 0  | 0  | 0  | 0  | 0  | 0  | 0   | 1   | 14  | 26  | 26  | 32  |
| GS  | 0  | 0  | 0  | 0  | 0  | 0  | 0  | 0  | 0   | 0   | 0   | 0   | 0   | 2   |
| GX  | 0  | 0  | 0  | 0  | 0  | 0  | 0  | 0  | 0   | 0   | 0   | 0   | 2   | 5   |
| GZ  | 0  | 0  | 0  | 0  | 0  | 0  | 0  | 0  | 0   | 0   | 0   | 0   | 1   | 3   |
| HA  | 0  | 0  | 0  | 0  | 0  | 0  | 0  | 0  | 0   | 0   | 0   | 1   | 5   | 5   |
| HB  | 41 | 41 | 41 | 41 | 41 | 41 | 45 | 62 | 121 | 198 | 270 | 375 | 444 | 549 |
| HE  | 0  | 0  | 0  | 0  | 0  | 0  | 0  | 0  | 0   | 0   | 0   | 0   | 1   | 1   |
| HI  | 0  | 0  | 0  | 0  | 0  | 0  | 0  | 0  | 0   | 0   | 0   | 0   | 4   | 5   |
| HL  | 0  | 0  | 0  | 0  | 0  | 0  | 0  | 0  | 0   | 0   | 0   | 0   | 0   | 2   |
| HN  | 0  | 0  | 0  | 0  | 0  | 0  | 0  | 0  | 0   | 0   | 0   | 1   | 4   | 9   |
| JL  | 0  | 0  | 0  | 0  | 0  | 0  | 0  | 0  | 0   | 0   | 0   | 0   | 0   | 1   |
| JS  | 0  | 0  | 0  | 0  | 0  | 0  | 0  | 0  | 0   | 0   | 0   | 0   | 1   | 5   |
| JX  | 0  | 0  | 0  | 0  | 0  | 0  | 0  | 0  | 0   | 0   | 0   | 2   | 2   | 7   |
| LN  | 0  | 0  | 0  | 0  | 0  | 0  | 0  | 0  | 0   | 0   | 0   | 0   | 2   | 3   |
| NM  | 0  | 0  | 0  | 0  | 0  | 0  | 0  | 0  | 0   | 0   | 0   | 0   | 0   | 0   |
| NX  | 0  | 0  | 0  | 0  | 0  | 0  | 0  | 0  | 0   | 0   | 0   | 0   | 1   | 1   |
| QH  | 0  | 0  | 0  | 0  | 0  | 0  | 0  | 0  | 0   | 0   | 0   | 0   | 0   | 0   |
| SC  | 0  | 0  | 0  | 0  | 0  | 0  | 0  | 0  | 0   | 0   | 0   | 2   | 5   | 8   |
| SD  | 0  | 0  | 0  | 0  | 0  | 0  | 0  | 0  | 0   | 0   | 0   | 1   | 2   | 6   |
| SH  | 0  | 0  | 0  | 0  | 0  | 0  | 0  | 0  | 0   | 0   | 1   | 6   | 9   | 16  |
| SN  | 0  | 0  | 0  | 0  | 0  | 0  | 0  | 0  | 0   | 0   | 0   | 0   | 0   | 3   |
| SX  | 0  | 0  | 0  | 0  | 0  | 0  | 0  | 0  | 0   | 0   | 0   | 0   | 1   | 1   |
| TJ  | 0  | 0  | 0  | 0  | 0  | 0  | 0  | 0  | 0   | 0   | 0   | 2   | 4   | 5   |
| XJ  | 0  | 0  | 0  | 0  | 0  | 0  | 0  | 0  | 0   | 0   | 0   | 0   | 0   | 2   |
| XZ  | 0  | 0  | 0  | 0  | 0  | 0  | 0  | 0  | 0   | 0   | 0   | 0   | 0   | 0   |
| YN  | 0  | 0  | 0  | 0  | 0  | 0  | 0  | 0  | 0   | 0   | 0   | 1   | 1   | 2   |
| ZJ  | 0  | 0  | 0  | 0  | 0  | 0  | 0  | 0  | 0   | 0   | 0   | 5   | 10  | 27  |
| SUM | 41 | 41 | 41 | 41 | 41 | 41 | 45 | 62 | 121 | 199 | 290 | 437 | 547 | 749 |

The short code for the provinces[7] can be obtained from Wikipedia (https://en.wikipedia.org/wiki/Provinces_of_China) or China Historical GIS. The data are collected from the CCDC[8] and CDCs of each province in mainland China.

**Table S1. Historical provincial data of 2019-nCoV confirmed patient amount from January 10 to February 5 (Part 2)**

|    | 24  | 25  | 26  | 27  | 28  | 29  | 30  | 31   | 01   | 02    | 03    | 04    | 05    |
|----|-----|-----|-----|-----|-----|-----|-----|------|------|-------|-------|-------|-------|
| AH | 15  | 39  | 60  | 70  | 106 | 152 | 200 | 237  | 297  | 340   | 408   | 480   | 530   |
| BJ | 36  | 41  | 68  | 80  | 91  | 111 | 121 | 139  | 168  | 191   | 212   | 228   | 253   |
| CQ | 27  | 57  | 75  | 110 | 132 | 165 | 206 | 238  | 262  | 275   | 337   | 366   | 389   |
| FJ | 10  | 18  | 35  | 59  | 80  | 84  | 101 | 120  | 144  | 159   | 179   | 194   | 205   |
| GD | 53  | 78  | 111 | 151 | 207 | 277 | 354 | 436  | 535  | 632   | 725   | 813   | 895   |
| GS | 4   | 4   | 7   | 14  | 19  | 24  | 29  | 35   | 40   | 40    | 55    | 57    | 62    |
| GX | 23  | 30  | 36  | 46  | 51  | 58  | 78  | 87   | 100  | 111   | 127   | 139   | 150   |
| GZ | 3   | 4   | 5   | 7   | 9   | 9   | 12  | 29   | 29   | 38    | 46    | 58    | 64    |
| HA | 9   | 32  | 83  | 128 | 168 | 206 | 278 | 352  | 422  | 493   | 566   | 675   | 764   |
| HB | 729 | 1052| 1423| 2714| 3554| 4586| 5806| 7153 | 9074 | 11177 | 13522 | 16678 | 19665 |
| HE | 2   | 8   | 13  | 18  | 33  | 48  | 65  | 82   | 96   | 104   | 113   | 126   | 135   |
| HI | 8   | 19  | 22  | 33  | 40  | 43  | 46  | 53   | 62   | 64    | 72    | 80    | 99    |
| HL | 4   | 9   | 16  | 21  | 33  | 38  | 44  | 59   | 80   | 95    | 121   | 155   | 190   |
| HN | 24  | 43  | 69  | 100 | 143 | 221 | 277 | 332  | 389  | 463   | 521   | 593   | 661   |
| JL | 3   | 4   | 5   | 6   | 8   | 9   | 14  | 14   | 17   | 23    | 31    | 42    | 54    |
| JS | 9   | 18  | 33  | 47  | 70  | 99  | 129 | 168  | 202  | 236   | 271   | 308   | 341   |
| JX | 18  | 30  | 48  | 72  | 109 | 162 | 162 | 240  | 286  | 333   | 391   | 476   | 548   |
| LN | 12  | 17  | 22  | 27  | 34  | 39  | 41  | 60   | 64   | 70    | 74    | 81    | 89    |
| NM | 2   | 7   | 7   | 11  | 15  | 16  | 19  | 20   | 23   | 27    | 34    | 35    | 42    |
| NX | 2   | 3   | 4   | 7   | 11  | 12  | 17  | 21   | 26   | 28    | 31    | 34    | 40    |
| QH | 0   | 1   | 3   | 6   | 6   | 7   | 8   | 8    | 9    | 11    | 13    | 15    | 17    |
| SC | 15  | 28  | 44  | 69  | 90  | 108 | 142 | 177  | 207  | 231   | 254   | 282   | 301   |
| SD | 15  | 27  | 46  | 75  | 95  | 130 | 158 | 184  | 206  | 230   | 259   | 275   | 307   |
| SH | 20  | 33  | 40  | 53  | 66  | 96  | 128 | 153  | 169  | 182   | 203   | 219   | 243   |
| SN | 5   | 15  | 22  | 35  | 46  | 56  | 63  | 87   | 101  | 116   | 128   | 142   | 165   |
| SX | 6   | 6   | 13  | 20  | 27  | 35  | 39  | 47   | 56   | 56    | 74    | 81    | 81    |
| TJ | 8   | 10  | 14  | 23  | 24  | 27  | 31  | 32   | 41   | 48    | 60    | 67    | 69    |
| XJ | 2   | 3   | 4   | 5   | 10  | 13  | 14  | 17   | 18   | 21    | 24    | 29    | 32    |
| XZ | 0   | 0   | 0   | 0   | 0   | 1   | 1   | 1    | 1    | 1     | 1     | 1     | 1     |
| YN | 5   | 11  | 16  | 26  | 44  | 55  | 76  | 83   | 93   | 105   | 117   | 122   | 128   |
| ZJ | 43  | 62  | 104 | 128 | 173 | 296 | 428 | 537  | 599  | 661   | 724   | 829   | 895   |
| SUM| 1112| 1709| 2448| 4161| 5494| 7183| 9087| 11201| 13816| 16561 | 19693 | 23680 | 27415 |

The short code for the provinces[7] can be obtained from Wikipedia (https://en.wikipedia.org/wiki/Provinces_of_China) or China Historical GIS. The data are collected from the CCDC[8] and CDCs of each province in mainland China.

**Table S2. Sigmoid function parameters for predicting 2019-nCoV transmission ending using Table S1**

| Province | *a* | *b* | *c* | *d* |
|---|---|---|---|---|
| AH | 362.075 | 0.170 | -24.029 | 357.012 |
| BJ | 171.578 | 0.127 | -22.858 | 165.851 |
| CQ | 225.083 | 0.167 | -21.613 | 218.971 |
| FJ | 115.760 | 0.183 | -21.414 | 112.812 |
| GD | 574.601 | 0.172 | -23.361 | 570.725 |
| GS | 36.712 | 0.176 | -22.280 | 35.898 |
| GX | 95.304 | 0.142 | -22.288 | 92.118 |
| GZ | 46.958 | 0.194 | -24.971 | 47.061 |
| HA | 518.483 | 0.168 | -24.041 | 509.693 |
| HB | 16544.009 | 0.157 | -25.921 | 16439.011 |
| HE | 69.305 | 0.248 | -21.343 | 68.521 |
| HI | 61.566 | 0.126 | -22.555 | 58.616 |
| HL | 356.310 | 0.136 | -30.689 | 354.362 |
| HN | 395.194 | 0.185 | -22.857 | 388.943 |
| JL | 1326.371 | 0.130 | -41.957 | 1326.191 |
| JS | 203.857 | 0.195 | -22.983 | 201.293 |
| JX | 423.549 | 0.157 | -25.130 | 417.565 |
| LN | 50.417 | 0.162 | -21.164 | 48.906 |
| NM | 30.129 | 0.130 | -24.047 | 29.086 |
| NX | 22.541 | 0.187 | -22.392 | 22.217 |
| QH | 11.900 | 0.133 | -23.776 | 11.437 |
| SC | 169.428 | 0.192 | -21.772 | 166.851 |
| SD | 164.396 | 0.195 | -21.291 | 160.804 |
| SH | 132.447 | 0.188 | -21.446 | 131.138 |
| SN | 100.129 | 0.169 | -22.859 | 97.998 |
| SX | 48.673 | 0.176 | -21.987 | 47.549 |
| TJ | 60.534 | 0.117 | -25.262 | 58.844 |
| XJ | 21.384 | 0.149 | -23.517 | 20.846 |
| XZ | 8.579 | 0.005 | -28.764 | 0.987 |
| YN | 66.187 | 0.237 | -20.708 | 65.263 |
| ZJ | 473.293 | 0.225 | -21.729 | 469.934 |
| SUM | 23373.179 | 0.149 | -25.858 | 23183.127 |

**Table S3. Gaussian function parameters for predicting 2019-nCoV transmission ending using Table S1**

| Province | *a* | *b* | *c* |
|---|---|---|---|
| AH | 60.282 | 24.464 | 3.905 |
| BJ | 21.802 | 23.999 | 5.325 |
| CQ | 35.329 | 21.648 | 3.967 |
| FJ | 19.675 | 21.073 | 3.512 |
| GD | 95.985 | 23.162 | 3.472 |
| GS | 6.078 | 22.315 | 3.773 |
| GX | 13.017 | 22.655 | 4.643 |
| GZ | 8.966 | 23.997 | 2.538 |
| HA | 90.518 | 25.817 | 4.506 |
| HB | 4247.754 | 32.126 | 5.548 |
| HE | 15.698 | 20.970 | 2.635 |
| HI | 28.500 | 45.626 | 10.441 |
| HL | 52.500 | 32.416 | 4.996 |
| HN | 69.933 | 23.349 | 3.777 |
| JL | 18.000 | 30.435 | 3.538 |
| JS | 37.724 | 23.077 | 3.398 |
| JX | 79.096 | 28.447 | 5.084 |
| LN | 7.653 | 21.447 | 4.219 |
| NM | 6.809 | 34.506 | 7.779 |
| NX | 4.112 | 24.134 | 4.348 |
| QH | 2.263 | 31.254 | 7.018 |
| SC | 30.576 | 21.475 | 3.265 |
| SD | 29.288 | 21.508 | 3.662 |
| SH | 23.241 | 21.551 | 3.535 |
| SN | 17.174 | 24.991 | 4.762 |
| SX | 8.081 | 21.284 | 3.353 |
| TJ | 6.416 | 23.915 | 4.693 |
| XJ | 3.445 | 26.351 | 5.333 |
| XZ | 1.000 | 19.000 | 0.174 |
| YN | 14.528 | 20.182 | 2.636 |
| ZJ | 98.858 | 21.546 | 2.890 |
| SUM | 4230.573 | 29.214 | 5.196 |

**Table S4. Poisson distribution parameters for predicting 2019-nCoV transmission ending using Table S1**

| Province | *a* | *b* |
|---|---|---|
| AH | 741.14 | 24.14 |
| BJ | 273.60 | 22.30 |
| CQ | 432.36 | 21.38 |
| FJ | 232.64 | 21.17 |
| GD | 1163.51 | 23.42 |
| GS | 73.89 | 22.18 |
| GX | 163.69 | 21.90 |
| GZ | 110.48 | 25.73 |
| HA | 1080.13 | 24.34 |
| HB | 37312.83 | 26.82 |
| HE | 166.55 | 21.68 |
| HI | 123.13 | 24.98 |
| HL | 505.63 | 28.81 |
| HN | 854.33 | 23.17 |
| JL | 295.66 | 32.11 |
| JS | 454.93 | 23.38 |
| JX | 855.91 | 25.32 |
| LN | 94.87 | 21.06 |
| NM | 53.28 | 24.26 |
| NX | 50.18 | 23.03 |
| QH | 20.94 | 23.85 |
| SC | 355.68 | 21.80 |
| SD | 351.84 | 21.44 |
| SH | 277.37 | 21.69 |
| SN | 206.98 | 23.23 |
| SX | 94.40 | 21.52 |
| TJ | 81.38 | 23.22 |
| XJ | 39.86 | 23.42 |
| XZ | 1.42 | 19.30 |
| YN | 150.56 | 20.72 |
| ZJ | 1108.19 | 22.19 |
| SUM | 44297.58 | 25.68 |

**Table S5. Moving-out transportation data of 2020 Chinese Spring Festival travel rush (Part 1)**

|    | AH    | BJ    | CQ    | FJ    | GD    | GS    | GX    | GZ    | HA    | HB    | HE    | HI    | HL    | HN    | JL    |
|----|-------|-------|-------|-------|-------|-------|-------|-------|-------|-------|-------|-------|-------|-------|-------|
| AH | 0     | 2E-02 | 8E-03 | 1E-02 | 2E-02 | 4E-03 | 5E-03 | 7E-03 | 1E-01 | 4E-02 | 2E-02 | 4E-03 | 4E-03 | 2E-02 | 3E-03 |
| BJ | 2E-02 | 0     | 8E-03 | 9E-03 | 3E-02 | 8E-03 | 5E-03 | 5E-03 | 7E-02 | 2E-02 | 4E-01 | 1E-02 | 2E-02 | 1E-02 | 1E-02 |
| CQ | 1E-02 | 2E-02 | 0     | 1E-02 | 4E-02 | 1E-02 | 1E-02 | 1E-01 | 2E-02 | 7E-02 | 1E-02 | 1E-02 | 3E-03 | 3E-02 | 2E-03 |
| FJ | 4E-02 | 2E-02 | 6E-02 | 0     | 1E-01 | 7E-03 | 3E-02 | 9E-02 | 5E-02 | 6E-02 | 1E-02 | 6E-03 | 5E-03 | 6E-02 | 4E-03 |
| GD | 2E-02 | 1E-02 | 4E-02 | 3E-02 | 0     | 3E-03 | 2E-01 | 6E-02 | 5E-02 | 9E-02 | 7E-03 | 1E-02 | 4E-03 | 2E-01 | 3E-03 |
| GS | 1E-02 | 3E-02 | 2E-02 | 1E-02 | 2E-02 | 0     | 8E-03 | 1E-02 | 5E-02 | 2E-02 | 2E-02 | 1E-02 | 4E-03 | 1E-02 | 3E-03 |
| GX | 1E-02 | 1E-02 | 3E-02 | 2E-02 | 4E-01 | 6E-03 | 0     | 9E-02 | 3E-02 | 3E-02 | 2E-02 | 2E-02 | 4E-03 | 1E-01 | 3E-03 |
| GZ | 1E-02 | 1E-02 | 2E-01 | 2E-02 | 8E-02 | 3E-03 | 7E-02 | 0     | 2E-02 | 3E-02 | 8E-03 | 1E-02 | 2E-03 | 1E-01 | 2E-03 |
| HA | 1E-01 | 6E-02 | 1E-02 | 2E-02 | 4E-02 | 1E-02 | 1E-02 | 9E-03 | 0     | 9E-02 | 8E-02 | 1E-02 | 5E-03 | 3E-02 | 4E-03 |
| HB | 6E-02 | 3E-02 | 7E-02 | 3E-02 | 8E-02 | 1E-02 | 2E-02 | 3E-02 | 1E-01 | 0     | 3E-02 | 9E-03 | 5E-03 | 1E-01 | 4E-03 |
| HE | 2E-02 | 4E-01 | 5E-03 | 4E-03 | 8E-03 | 4E-03 | 3E-03 | 3E-03 | 7E-02 | 2E-02 | 0     | 4E-03 | 2E-02 | 8E-03 | 1E-02 |
| HI | 2E-02 | 5E-02 | 4E-02 | 2E-02 | 2E-01 | 1E-02 | 6E-02 | 4E-02 | 5E-02 | 4E-02 | 3E-02 | 0     | 2E-02 | 6E-02 | 1E-02 |
| HL | 2E-02 | 9E-02 | 7E-03 | 1E-02 | 3E-02 | 7E-03 | 7E-03 | 1E-02 | 3E-02 | 1E-02 | 5E-02 | 2E-02 | 0     | 1E-02 | 2E-01 |
| HN | 2E-02 | 3E-02 | 4E-02 | 2E-02 | 2E-01 | 8E-03 | 7E-02 | 7E-02 | 4E-02 | 1E-01 | 2E-02 | 1E-02 | 5E-03 | 0     | 4E-03 |
| JL | 2E-02 | 7E-02 | 6E-03 | 7E-03 | 2E-02 | 6E-03 | 6E-03 | 7E-03 | 3E-02 | 1E-02 | 5E-02 | 1E-02 | 2E-01 | 1E-02 | 0     |
| JS | 2E-01 | 2E-02 | 1E-02 | 1E-02 | 2E-02 | 1E-02 | 6E-03 | 2E-02 | 9E-02 | 3E-02 | 2E-02 | 3E-03 | 6E-03 | 2E-02 | 4E-03 |
| JX | 6E-02 | 2E-02 | 2E-02 | 8E-02 | 2E-01 | 1E-02 | 2E-02 | 3E-02 | 5E-02 | 8E-02 | 2E-02 | 9E-03 | 6E-03 | 1E-01 | 5E-03 |
| LN | 2E-02 | 1E-01 | 7E-03 | 1E-02 | 2E-02 | 6E-03 | 7E-03 | 9E-03 | 4E-02 | 1E-02 | 1E-01 | 1E-02 | 1E-01 | 9E-03 | 2E-01 |
| NM | 8E-03 | 1E-01 | 4E-03 | 5E-03 | 1E-02 | 3E-02 | 2E-03 | 2E-03 | 2E-02 | 7E-03 | 1E-01 | 8E-03 | 1E-01 | 4E-03 | 7E-02 |
| NX | 1E-02 | 3E-02 | 1E-02 | 9E-03 | 2E-02 | 2E-01 | 9E-03 | 4E-03 | 4E-02 | 2E-02 | 3E-02 | 7E-03 | 4E-03 | 1E-02 | 3E-03 |
| QH | 2E-02 | 2E-02 | 2E-02 | 8E-03 | 2E-02 | 4E-01 | 3E-03 | 5E-03 | 7E-02 | 2E-02 | 2E-02 | 9E-03 | 3E-03 | 1E-02 | 1E-03 |
| SC | 1E-02 | 3E-02 | 3E-01 | 1E-02 | 5E-02 | 3E-02 | 2E-02 | 8E-02 | 3E-02 | 3E-02 | 2E-02 | 1E-02 | 5E-03 | 2E-02 | 4E-03 |
| SD | 5E-02 | 8E-02 | 9E-03 | 1E-02 | 2E-02 | 8E-03 | 7E-03 | 9E-03 | 1E-01 | 3E-02 | 1E-01 | 6E-03 | 3E-02 | 1E-02 | 2E-02 |
| SH | 9E-02 | 3E-02 | 2E-02 | 2E-02 | 3E-02 | 7E-03 | 6E-03 | 1E-02 | 5E-02 | 2E-02 | 1E-02 | 6E-03 | 7E-03 | 2E-02 | 5E-03 |
| SN | 3E-02 | 4E-02 | 3E-02 | 1E-02 | 3E-02 | 1E-01 | 9E-03 | 1E-02 | 1E-01 | 5E-02 | 4E-02 | 1E-02 | 6E-03 | 2E-02 | 5E-03 |
| SX | 2E-02 | 9E-02 | 1E-02 | 1E-02 | 2E-02 | 8E-03 | 6E-03 | 5E-03 | 1E-01 | 2E-02 | 2E-01 | 1E-02 | 5E-03 | 1E-02 | 5E-03 |
| TJ | 3E-02 | 2E-01 | 7E-03 | 7E-03 | 1E-02 | 1E-02 | 7E-03 | 9E-03 | 5E-02 | 1E-02 | 4E-01 | 6E-03 | 2E-02 | 8E-03 | 1E-02 |
| XJ | 2E-02 | 4E-02 | 3E-02 | 9E-03 | 3E-02 | 3E-01 | 6E-03 | 5E-03 | 1E-01 | 2E-02 | 3E-02 | 1E-02 | 5E-03 | 1E-02 | 4E-03 |
| XZ | 1E-02 | 9E-03 | 6E-02 | 5E-03 | 8E-03 | 5E-02 | 2E-03 | 9E-03 | 4E-02 | 2E-02 | 1E-02 | 2E-03 | 2E-03 | 1E-02 | 6E-04 |
| YN | 1E-02 | 2E-02 | 8E-02 | 2E-02 | 5E-02 | 6E-03 | 6E-02 | 2E-01 | 3E-02 | 3E-02 | 2E-02 | 7E-03 | 5E-03 | 4E-02 | 4E-03 |
| ZJ | 1E-01 | 1E-02 | 4E-02 | 3E-02 | 2E-02 | 5E-03 | 1E-02 | 1E-01 | 9E-02 | 6E-02 | 9E-03 | 3E-03 | 4E-03 | 6E-02 | 3E-03 |

The data in (row A, column B) mean the ratios between the people departing from A to B and the moving-out people in A. The "E" in each item denotes the base of 10, and the "$a\mathrm{E}b$" means $a \times 10^{b}$. For example, 2E-02 denotes $2 \times 10^{-2}$. The transportation data are collected from Baidu.

**Table S5. Moving-out transportation data of 2020 Chinese Spring Festival travel rush (Part 2)**

|    | JS    | JX    | LN    | NM    | NX    | QH    | SC    | SD    | SH    | SN    | SX    | TJ    | XJ    | XZ    | YN    | ZJ    |
|----|-------|-------|-------|-------|-------|-------|-------|-------|-------|-------|-------|-------|-------|-------|-------|-------|
| AH | 4E-01 | 3E-02 | 5E-03 | 2E-03 | 1E-03 | 7E-04 | 1E-02 | 5E-02 | 8E-02 | 1E-02 | 7E-03 | 6E-03 | 2E-03 | 3E-04 | 6E-03 | 1E-01 |
| BJ | 3E-02 | 1E-02 | 3E-02 | 3E-02 | 3E-03 | 1E-03 | 2E-02 | 6E-02 | 2E-02 | 2E-02 | 4E-02 | 7E-02 | 4E-03 | 3E-04 | 7E-03 | 2E-02 |
| CQ | 2E-02 | 1E-02 | 5E-03 | 3E-03 | 3E-03 | 3E-03 | 5E-01 | 1E-02 | 1E-02 | 2E-02 | 7E-03 | 5E-03 | 7E-03 | 2E-03 | 4E-02 | 2E-02 |
| FJ | 3E-02 | 1E-01 | 6E-03 | 3E-03 | 1E-03 | 6E-04 | 6E-02 | 2E-02 | 3E-02 | 1E-02 | 8E-03 | 5E-03 | 2E-03 | 3E-04 | 3E-02 | 9E-02 |
| GD | 2E-02 | 9E-02 | 5E-03 | 2E-03 | 7E-04 | 4E-04 | 7E-02 | 1E-02 | 1E-02 | 1E-02 | 4E-03 | 3E-03 | 2E-03 | 1E-04 | 3E-02 | 2E-02 |
| GS | 4E-02 | 8E-03 | 6E-03 | 3E-02 | 9E-02 | 9E-02 | 9E-02 | 2E-02 | 2E-02 | 3E-01 | 2E-02 | 6E-03 | 5E-02 | 4E-03 | 1E-02 | 2E-02 |
| GX | 2E-02 | 3E-02 | 7E-03 | 3E-03 | 2E-03 | 1E-03 | 4E-02 | 2E-02 | 9E-03 | 1E-02 | 7E-03 | 4E-03 | 1E-03 | 1E-04 | 7E-02 | 2E-02 |
| GZ | 2E-02 | 2E-02 | 3E-03 | 1E-03 | 4E-04 | 4E-04 | 2E-01 | 9E-03 | 9E-03 | 1E-02 | 4E-03 | 2E-03 | 7E-04 | 1E-04 | 2E-01 | 3E-02 |
| HA | 9E-02 | 2E-02 | 9E-03 | 6E-03 | 4E-03 | 3E-03 | 2E-02 | 1E-01 | 3E-02 | 6E-02 | 6E-02 | 2E-02 | 6E-03 | 3E-04 | 9E-03 | 4E-02 |
| HB | 5E-02 | 7E-02 | 8E-03 | 6E-03 | 3E-03 | 3E-03 | 4E-02 | 3E-02 | 2E-02 | 3E-02 | 2E-02 | 6E-03 | 6E-03 | 7E-04 | 2E-02 | 4E-02 |
| HE | 2E-02 | 6E-03 | 3E-02 | 3E-02 | 2E-03 | 8E-04 | 1E-02 | 1E-01 | 6E-03 | 2E-02 | 6E-02 | 1E-01 | 2E-03 | 1E-04 | 4E-03 | 1E-02 |
| HI | 3E-02 | 3E-02 | 2E-02 | 8E-03 | 3E-03 | 2E-03 | 7E-02 | 3E-02 | 3E-02 | 3E-02 | 2E-02 | 2E-02 | 7E-03 | 2E-04 | 2E-02 | 3E-02 |
| HL | 3E-02 | 8E-03 | 1E-01 | 1E-01 | 2E-03 | 1E-03 | 2E-02 | 6E-02 | 2E-02 | 1E-02 | 2E-02 | 3E-02 | 3E-03 | 2E-04 | 8E-03 | 2E-02 |
| HN | 3E-02 | 9E-02 | 6E-03 | 4E-03 | 2E-03 | 2E-03 | 3E-02 | 2E-02 | 2E-02 | 2E-02 | 8E-03 | 4E-03 | 5E-03 | 1E-03 | 2E-02 | 3E-02 |
| JL | 3E-02 | 7E-03 | 2E-01 | 1E-01 | 3E-03 | 1E-03 | 1E-02 | 5E-02 | 2E-02 | 1E-02 | 2E-02 | 2E-02 | 4E-03 | 3E-04 | 6E-03 | 2E-02 |
| JS | 0     | 2E-02 | 8E-03 | 3E-03 | 2E-03 | 2E-03 | 2E-02 | 9E-02 | 2E-01 | 2E-02 | 1E-02 | 6E-03 | 3E-03 | 5E-04 | 1E-02 | 1E-01 |
| JX | 5E-02 | 0     | 7E-03 | 4E-03 | 2E-03 | 2E-03 | 2E-02 | 2E-02 | 4E-02 | 1E-02 | 9E-03 | 5E-03 | 3E-03 | 6E-04 | 1E-02 | 1E-01 |
| LN | 4E-02 | 7E-03 | 0     | 1E-01 | 2E-03 | 2E-03 | 2E-02 | 7E-02 | 2E-02 | 1E-02 | 2E-02 | 3E-02 | 5E-03 | 4E-04 | 8E-03 | 2E-02 |
| NM | 1E-02 | 3E-03 | 1E-01 | 0     | 8E-02 | 2E-03 | 9E-03 | 3E-02 | 7E-03 | 1E-01 | 8E-02 | 2E-02 | 2E-03 | 0E+00 | 4E-03 | 7E-03 |
| NX | 3E-02 | 6E-03 | 6E-03 | 2E-01 | 0     | 1E-02 | 2E-02 | 2E-02 | 2E-02 | 2E-01 | 2E-02 | 5E-03 | 1E-02 | 6E-04 | 1E-02 | 2E-02 |
| QH | 2E-02 | 4E-03 | 3E-03 | 9E-03 | 2E-02 | 0     | 9E-02 | 2E-02 | 5E-03 | 1E-01 | 1E-02 | 3E-03 | 2E-02 | 4E-02 | 9E-03 | 1E-02 |
| SC | 2E-02 | 1E-02 | 7E-03 | 5E-03 | 4E-03 | 7E-03 | 0     | 2E-02 | 2E-02 | 6E-02 | 1E-02 | 5E-03 | 1E-02 | 8E-03 | 1E-01 | 2E-02 |
| SD | 2E-01 | 1E-02 | 4E-02 | 1E-02 | 3E-03 | 2E-03 | 3E-02 | 0     | 3E-02 | 2E-02 | 3E-02 | 4E-02 | 4E-03 | 1E-04 | 9E-03 | 3E-02 |
| SH | 4E-01 | 3E-02 | 9E-03 | 3E-03 | 1E-03 | 6E-04 | 2E-02 | 3E-02 | 0     | 1E-02 | 8E-03 | 6E-03 | 3E-03 | 2E-04 | 1E-02 | 2E-01 |
| SN | 4E-02 | 1E-02 | 9E-03 | 6E-02 | 4E-02 | 1E-02 | 1E-01 | 4E-02 | 2E-02 | 0     | 1E-01 | 8E-03 | 2E-02 | 3E-03 | 1E-02 | 2E-02 |
| SX | 3E-02 | 8E-03 | 1E-02 | 8E-02 | 6E-03 | 3E-03 | 3E-02 | 6E-02 | 1E-02 | 1E-01 | 0     | 3E-02 | 2E-03 | 2E-04 | 9E-03 | 2E-02 |
| TJ | 2E-02 | 6E-03 | 3E-02 | 2E-02 | 3E-03 | 2E-03 | 1E-02 | 8E-02 | 1E-02 | 1E-02 | 4E-02 | 0     | 5E-03 | 4E-04 | 7E-03 | 1E-02 |
| XJ | 3E-02 | 5E-03 | 8E-03 | 2E-02 | 2E-02 | 3E-02 | 1E-01 | 3E-02 | 1E-02 | 8E-02 | 1E-02 | 6E-03 | 0     | 1E-03 | 9E-03 | 2E-02 |
| XZ | 9E-03 | 5E-03 | 3E-03 | 9E-04 | 3E-03 | 1E-01 | 4E-01 | 1E-02 | 4E-03 | 6E-02 | 6E-03 | 2E-03 | 6E-03 | 0     | 5E-02 | 6E-03 |
| YN | 2E-02 | 2E-02 | 7E-03 | 4E-03 | 2E-03 | 1E-03 | 3E-01 | 2E-02 | 1E-02 | 2E-02 | 8E-03 | 5E-03 | 3E-03 | 2E-03 | 0     | 3E-02 |
| ZJ | 1E-01 | 1E-01 | 5E-03 | 2E-03 | 8E-04 | 3E-04 | 4E-02 | 2E-02 | 9E-02 | 1E-02 | 5E-03 | 2E-03 | 1E-03 | 0E+00 | 3E-02 | 0     |

The data in (row A, column B) mean the ratios between the people departing from A to B and the moving-out people in A. The "E" in each item denotes the base of 10, and the "$a\mathrm{E}b$" means $a \times 10^b$. For example, 2E-02 denotes $2 \times 10^{-2}$. The transportation data are collected from Baidu.

**Table S6. Training results of SEIRSD MMODEs-NN**

| | $\alpha$ | $\beta$ | $\gamma_1$ | $\gamma_2$ | $\delta$ | $\theta$ | $P_C$ | Hubei $P_C$ | Total confirmed MAE |
|---|---|---|---|---|---|---|---|---|---|
| **No delay** | 1.00 | 0.58 | 0.39 | 0.01 | 0.009 | 0.00 | 0.0008 | 0.0864 | 1611.74 |
| **One-day** | 1.00 | 0.59 | 0.41 | 0.01 | 0.009 | 0.00 | 0.0008 | 0.1032 | 1296.33 |
| **Two-day** | 1.00 | 0.60 | 0.38 | 0.01 | 0.009 | 0.00 | 0.0008 | 0.1352 | 1331.33 |
| **No transportation limitation policies** | 1.00 | 0.80 | 0.30 | 0.01 | 0.100 | 0.00 | 0.0008 | 0.0864 | 9867.48 |

**Table S7. Total Confirmed amount prediction results of SEIRSD MMODEs-NN Simulations in mainland China**

| Date | No delay | One-day delay | Two-day delay | No transportation limitation policies | Reality |
|---|---|---|---|---|---|
| 2020/2/1 | 16175 | 14866 | 14247 | 36915 | 13816 |
| 2020/2/2 | 19093 | 17947 | 17658 | 44973 | 16561 |
| 2020/2/3 | 22141 | 21308 | 21566 | 54092 | 19693 |
| 2020/2/4 | 25168 | 24830 | 25836 | 64161 | 23680 |
| 2020/2/5 | 28036 | 28318 | 30402 | 74857 | 27415 |
| 2020/2/6 | 30638 | 31652 | 34970 | 86275 | 31211 |
| 2020/2/7 | 32908 | 34661 | 39423 | 98637 | 34598 |
| 2020/2/8 | 34763 | 37249 | 43461 | 111145 | 37198 |
| 2020/2/9 | 36251 | 39402 | 46974 | 124667 | 40171 |
| 2020/2/10 | 37408 | 41101 | 49863 | 139896 | 42638 |
| 2020/2/11 | 38284 | 42426 | 52191 | 154289 | 44653 |
| 2020/2/12 | 38928 | 43416 | 53964 | 170937 | 46551 (59883) |
| 2020/2/13 | 39395 | 44132 | 55290 | 188166 | 48288 (63851) |
| 2020/2/14 | 39735 | 44666 | 56265 | 206676 | 49791 (66576) |
| 2020/2/15 | 39976 | 45042 | 56961 | 224008 | 50912 (68584) |
| 2020/2/16 | 40150 | 45316 | 57461 | 242632 | 52960 (70635) |
| 2020/2/17 | 40270 | 45504 | 57808 | 262303 | 54846 (72528) |
| 2020/2/18 | 40356 | 45639 | 58048 | 279664 | 56595 (74279) |
| 2020/2/19 | 40415 | 45732 | 58215 | 298669 | 56989 (74675) |
| 2020/2/20 | 40457 | 45797 | 58328 | 315552 | -- |
| 2020/2/21 | 40484 | 45841 | 58405 | 332321 | -- |
| 2020/2/22 | 40505 | 45873 | 58462 | 347140 | -- |
| 2020/2/23 | 40517 | 45893 | 58495 | 361771 | -- |
| 2020/2/24 | 40525 | 45905 | 58517 | 373450 | -- |
| 2020/2/25 | 40530 | 45914 | 58533 | 384479 | -- |
| 2020/2/26 | 40530 | 45920 | 58540 | 394220 | -- |
| 2020/2/27 | 40530 | 45926 | 58545 | 401930 | -- |
| 2020/2/28 | 40532 | 45926 | 58546 | 408992 | -- |
| 2020/2/29 | 40533 | 45926 | 58547 | 415307 | -- |
| 2020/3/1 | 40533 | 45926 | 58547 | 420487 | -- |
| 2020/3/2 | 40533 | 45926 | 58547 | 424979 | -- |
| 2020/3/3 | 40533 | 45926 | 58547 | 429093 | -- |
| 2020/3/4 | 40533 | 45926 | 58547 | 432650 | -- |
| 2020/3/5 | 40533 | 45926 | 58547 | 435338 | -- |
| 2020/3/6 | 40533 | 45926 | 58547 | 437927 | -- |
| 2020/3/7 | 40533 | 45926 | 58547 | 440011 | -- |
| 2020/3/8 | 40533 | 45926 | 58547 | 441960 | -- |
| 2020/3/9 | 40533 | 45926 | 58547 | 443753 | -- |

**The CDC of China published a lower standard to judge the confirmed infected patients on February 9, 2020. On February 12, Hubei Province started to use the new standard to calculate the confirmed patient count. The numbers in brackets are the statistic results under the new standard.**

**References**

1. Li, M. and Muldowney, J. Global stability for the SEIR model in epidemiology. *Mathematical biosciences* **125**, 2, 155-164, (1995).
2. Li, J., Ye, Q., Deng, X., Liu, Y., and Liu, Y. Spatial-temporal analysis on Spring Festival travel rush in China based on multisource big data. *Sustainability* **8**, 11, 1184, (2016).
3. Zhang, Y. *Analysis and Design of Recurrent Neural Networks and Their Applications to Control and Robotic Systems*, Ph.D. Thesis, Chinese University of Hong Kong, Hong Kong, (2002).
4. Williams, R., Hinton, G., and Rumelhart, D. Learning representations by back-propagating errors. *Nature* **323**, 6088, 533-536, (1986).
5. Zhang, Y., Guo, D., and Li, Z. Common nature of learning between back-propagation and Hopfield-type neural networks for generalized matrix inversion with simplified models. *IEEE Transactions on Neural Networks and Learning Systems* **24**, 4, 579-592, (2013).
6. Moller, M. *A Scaled Conjugate Gradient Algorithm for Fast Supervised Learning*, Aarhus University, Aarhus, Denmark, (1990).
7. Bol, P. and Ge, J. China historical GIS. *Historical Geography* **33**, 150-152, (2005).
8. CCDC, *Epidemic Update and Risk Assessment of 2019 Novel Coronavirus*, CCDC, Beijing, China, http://www.chinacdc.cn/yyrdgz/202001/P020200128523354919292.pdf, (in Chinese, 2020).